\documentclass[epj]{svjour}
\usepackage{times}
\usepackage{graphicx}
\usepackage{amsmath}
\usepackage{amssymb}

\newcommand{\ud}{\mathrm{d}}
\newcommand{\dasgraph}[1]{\centerline{\resizebox{0.75\columnwidth}{!}{\includegraphics{#1}}}}

\begin{document}

\title{Coherence length of  an
elongated condensate: a study by matter-wave interferometry}
\subtitle{\\}

\author{M. Hugbart\inst{1}\thanks{e-mail: mathilde.hugbart@iota.u-psud.fr},
J. A. Retter\inst{1}, F. Gerbier\inst{2}, A. Varon\inst{1}, S.
Richard\inst{3}, J. H. Thywissen\inst{4}, D. Clement\inst{1}, P.
Bouyer\inst{1}, A. Aspect\inst{1}}

\institute{Laboratoire Charles Fabry de l'Institut d'Optique,
UMR8501 du CNRS, 91403 Orsay Cedex, France \and Johannes
Gutenberg-Universit\"{a}t, 55099 Mainz, Germany \and THALES
Research \& Technology, 91404 Orsay Cedex, France \and Department
of Physics, University of Toronto, Canada}

\date{Received: date / Revised version: date}

\abstract{ We measure the spatial correlation function of
Bose-Einstein condensates in the cross-over region between
phase-coherent and strongly phase-fluctuating condensates. We
observe the continuous path from a gaussian-like shape to an
exponential-like shape characteristic of one-dimensional
phase-fluctuations. The width of the spatial correlation function
as a function of the temperature shows that the condensate
coherence length undergoes no sharp transition between these two
regimes.  
}

\authorrunning{M. Hugbart {\it et al.}}

\maketitle


\section{Introduction}

The high atomic phase-space density provided by a
Bose-\linebreak[4]Einstein condensate (BEC) \cite{BEC} has driven
interest in guiding atoms in a manner analogous to guiding laser
light through single-mode optical fibres. The most advanced
technology to date is based on \textit{atom chips}, where the
fields which trap and guide the atoms are created by
microfabricated structures
\cite{ReiHan99,Cor99,Dek00,Smi01,Zim02,Hin03}. Many groups around
the world have already succeeded in preparing BEC on an atom chip
\cite{Han01,Ott01,Gus02,Sch03,Vul04,Est04,Val04} and one can
envisage using this technology to create integrated atom
interferometers \cite{Hin01,HanA01,And02,Wang04}. In this context,
a precise characterization of the phase coherence properties of
the condensate is crucial \cite{Det01,Shv02,Ric03}.

Trapping quantum gases on atom chips naturally involves highly
elongated, guide-like traps. Changing the dimensionality of the
system from three-dimensional (3D) towards one-dimensional (1D)
has a profound effect on the phase coherence of the condensate. In
3D condensates of modest aspect ratio, experimental results show
that phase coherence extends across the whole cloud
\cite{Hag99,Ste99}, even at finite temperature \cite{Blo00}.
However, in the 1D regime, thermal excitations of the low energy
axial modes lead to phase fluctuations which degrade the phase
coherence \cite{Pet00,Cas00}. Condensates exhibiting such phase
fluctuations are known as {\it quasi-condensates}.

In the intermediate regime of elongated 3D condensates with a high
aspect ratio, a behaviour similar to the 1D case is observed
\cite{Pet01}: below a characteristic temperature $T_{\phi}$
determined by the atom number and the trapping frequencies, the
condensate is nearly phase coherent, but above $T_{\phi}$ the
population of the axial modes is high and phase fluctuations may
be pronounced.  For weakly elongated condensates, $T_\phi$ can be
higher than the transition temperature $T_{\rm{c}}$, so that the
condensate is nearly  phase coherent at all temperatures. In
contrast, atom chips can easily produce traps with high  aspect
ratios ($\sim\!1000$) for which $T_\phi$ can be much smaller than
$T_{\rm{c}}$. Phase fluctuations are therefore likely to impose
limits on the performance of atom chip devices and need to be well
understood.

In an elongated condensate, the wavelength of the low energy axial
modes is longer than the radial size of the condensate, so these
excitations have a 1D character. However, the wavelength of these
excitations can be much shorter than the axial length of the
condensate, reducing the phase coherence length in the axial
direction. An important feature of quasi-condensates is that
density fluctuations remain suppressed in the trap, due to the
mean field energy, even in the presence of large phase
fluctuations \cite{Pet01,Ric03}. Therefore, the 3D
quasi-condensate has the usual parabolic profile in the
Thomas-Fermi limit.

Phase-fluctuating condensates in elongated traps were first
observed by the conversion of phase fluctuations into density
fluctuations after a sufficiently long free expansion \cite{Det01}
and by a condensate-focussing technique \cite{Shv02}. Quantitative
measurements of the phase coherence length have since been
obtained from the momentum distribution \cite{Ric03} and the
second-order correlation function \cite{Hel03}. The results of
each of these experiments showed good agreement with theory
\cite{Pet01} in the strongly phase-fluctuating regime (at
temperatures $T\gg T_{\phi}$). However, neither experiment
explored the cross-over region ($T\sim T_\phi$) between
 phase-coherent and strongly phase-fluctuating condensates.

In this article we describe a new experiment using a matter-wave
interferometer and Fourier-space analysis to measure the spatial
correlation function, thereby extending our measurements into the
cross-over region $(T\sim T_{\phi})$. Our results agree with the
predicted shapes of the correlation function: for $T\gg T_{\phi}$,
we find exponential-like correlation functions as predicted for
significant phase fluctuations, whereas at $T\sim T_\phi$ we find
a gaussian-like shape, as expected when the phase profile is
nearly flat and the correlation function decay is dominated by the
density profile. The coherence length as a function of
$T/T_{\phi}$ follows the trend predicted by theory, showing that
the coherence length increases smoothly as the temperature falls
and that there is no sharp transition at $T_{\phi}$. This
highlights the fact that phase fluctuations occur at all finite
temperatures, even if these effects are too small to be resolved
experimentally for more spherical traps. However, whereas our
previous measurements based on momentum spectroscopy \cite{Ric03},
realized for high $T/T_\phi$, were in full agrement with the
theory, two observations remain unexplained in the interferometric
method. First, as in a previous experiment \cite{Hag99}, our
experimental measurements of the coherence length are shifted from
the theoretical prediction, by about $20\%$ for $T/T_\phi=0$, even
after taking the limitations of our imaging system into account.
Second, our interferometer produces unexplained supplementary
fringes outside of the region where the condensates overlap, and
we note that similar unexplained fringes appear in other published
data \cite{Sim00}. These supplementary fringes do not seem to be
compatible with interference of the thermal cloud observed in
\cite{KetterleThermal}.

\section{Measurement of the coherence length by atom interferometry}
A natural method to study the coherence length along the long axis
$z$ of a condensate, or a quasi-condensate, is to use atom
interferometry. With atomic beam-splitters, one produces two
daughter copies of the initial condensate with a separation $s$,
and observes the interference pattern appearing in the atomic
density:
\begin{eqnarray}
n(\Vec{r}) &\propto& |\Psi_0(\vec{r}-\frac {s}{2}\vec{e}_z)
+e^{i\phi_{\rm{rel}}}\Psi_0(\vec{r}+\frac {s}{2}\vec{e}_z)
|^2 \nonumber\\
&\propto& |\Psi_0(\vec{r}-\frac {s}{2}\vec{e}_z)|^2
+|\Psi_0(\vec{r}+\frac
{s}{2}\vec{e}_z) |^2 \nonumber\\
&+& 2{\rm Re}[e^{i\phi_{\rm{rel}}}\Psi_0^{\ast}(\vec{r}-\frac
{s}{2}\vec{e}_z)\Psi_0(\vec{r}+\frac {s}{2}\vec{e}_z)],
\label{equ:interfometer}
\end{eqnarray}
where $\vec{e}_z$ is the axial unit vector, $\Psi_0(\vec{r})$ is
the wavefunction describing the initial condensate and
$\phi_{\rm{rel}}$ a relative phase shift produced by the
interferometer and the free fall of the condensate.

Let us first consider the behaviour of a fully phase coherent
condensate. During free expansion, it acquires a phase
distribution proportional to $z^2$ \cite{Cas96}. The phase
difference $\phi_{\rm{rel}}$ between two displaced copies of the
condensate is therefore proportional to $zs$, giving rise to an
interference pattern of straight fringes, uniformly spaced along
the longitudinal direction, the spatial frequency of the fringes
being proportional to the separation $s$. The fringe contrast
integrated over the entire condensate gives the first-order
correlation function at $s$:
\begin{eqnarray}\label{equ:C1}
C^{(1)}(s) = \int \ud^3 \vec{r}\,\Psi_0^{\ast}(\vec{r}-\frac
{s}{2}\vec{e}_z)\Psi_0(\vec{r}+\frac {s}{2}\vec{e}_z).
\end{eqnarray}
Repeating this contrast measurement for different separations $s$,
one can study the decay of $C^{(1)}(s)$ with increasing $s$
\cite{Hag99}.  For a fully phase coherent condensate, the
first-order correlation function decay reflects only the width of
the density profile $n(\vec{r})$ \cite{Zam00}.

In the case of a quasi-condensate, 1D thermal excitations cause
the phase to fluctuate along the longitudinal axis, both spatially
and temporally.  In our experiment, these fluctuations are small
compared to the parabolic phase developed during free expansion.
Therefore when we image the overlapping condensates after free
expansion, we still observe straight fringes, but they are no
longer strictly periodic.  Small local phase shifts add a
``jitter'' to the fringe spacing, which in Fourier space has the
effect of broadening the peak at the spatial frequency of the
fringes and thereby reducing its height. As $s$ is increased, the
fringes are perturbed more strongly, because the condensate phase
becomes less correlated at larger separations. The contrast
therefore decreases faster with $s$ than in the fully
phase-coherent case. The greater the amplitude of the phase
fluctuations, the faster the contrast decreases with $s$.
Therefore by measuring the width of the correlation function at
different temperatures, we extract the temperature dependence of
the coherence length.  Further information is obtained from the
shape of the correlation function \cite{Ger03}.

In the presence of phase fluctuations, each realization of the
experiment gives a different interference pattern, even with fixed
experimental conditions.   Expression (\ref{equ:C1}) must
therefore be generalized to:
\begin{eqnarray}\label{equ:Cmean}
C^{(1)}(s) = \int \ud^3 \vec{r}\,\langle
\Psi_0^{\ast}(\vec{r}-\frac {s}{2}\vec{e}_z)\Psi_0(\vec{r}+\frac
{s}{2}\vec{e}_z)\rangle,
\end{eqnarray}
where the brackets $\langle\rangle$ denote a statistical average
of the random process describing the random phase. In practice,
one must repeat the experiment at a given separation $s$ and
average the contrast measurements over many quasi-condensates to
obtain $C^{(1)}(s)$.

This principle was  used by Michelson in his famous astronomical
interferometer, whose goal was to measure the spatial coherence of
the light field arriving from a star, in order to deduce the
diameter of the star \cite{Michelson}. However, Michelson's method
is plagued by the existence of a randomly fluctuating relative
phase between the two inputs of the interferometer, and various
methods insensitive to the relative phase fluctuations had to be
developed \cite{Labeyrie}. A similar problem appears when one
tries to determine the coherence length of a condensate or a
quasi-condensate with interferometry. In addition to the
controlled relative phase between the two components interfering
in (\ref{equ:interfometer}), there is an uncontrollable relative
phase due to experimental problems such as a residual phase shift
between the lasers creating the two copies or a random velocity
kick imparted to the sample. In order to overcome this problem
\cite{Bragg}, D. Hellweg \textit{et al.} \cite{Hel03} have used an
analysis analogous to the Hanbury Brown and Twiss method
\cite{HBT}, since it is based on the measurement of the
second-order correlation function which is insensitive to  global
phase shifts. In contrast, our method is in line with the initial
method of Michelson who could visually evaluate the contrast of
the randomly moving fringes he was observing. The decrease of this
contrast as a function of the telescopes' separation gave a direct
measurement of the coherence length. Similarly, we directly
evaluate the contrast of the fringes by taking the modulus of the
Fourier transform of the fringe pattern. The decrease of that
contrast as a function of the separation $s$ yields the coherence
length of the quasi-condensate.

\section{Experiment}

\subsection{Creation of elongated Bose-Einstein condensates}
In our experimental setup \cite{Des99}, a Zeeman-slowed atomic
beam of $^{87}$Rb is trapped in a MOT, and after optical pumping
into the $5S_{1/2}|F=1,m_F=-1\rangle$ state is transferred to a
magnetic Ioffe-Pritchard trap created by an iron-core
electromagnet.  Our design allows us to lower the magnetic field
at the bottom of the trap to a few Gauss and thus to obtain very
tight radial confinement \cite{Bou00}. Using this trap, we are
able to create condensates very close to the 1D Thomas-Fermi
regime \cite{Str98}, as was demonstrated in
\cite{QGLD,RicT03,GerT03}. In the present experiment, we produce
condensates further into the 3D regime so that we can explore the
cross-over regime ($T\sim T_{\phi}$). We use two different trap
configurations:  in the first, the final radial and axial trap
frequencies are respectively $\omega_\bot=2\pi \times 395\,$Hz and
$\omega_z=2\pi \times 8.67\,$Hz, giving an aspect ratio of 45; for
the second trap configuration, the final frequencies are
$\omega_\bot=2\pi \times 655\,$Hz and $\omega_z=2\pi \times
6.55\,$Hz, with aspect ratio 100.  In this way, we obtain
needle-shaped condensates containing around $3\times 10^5$\,atoms,
with a typical half-length $L\,\simeq\,85\,\mu$m in the first trap
and $L\,\simeq\,120\,\mu$m in the second. We control the final
number of atoms by holding the condensate for a variable time,
typically a few seconds, in the presence of an rf shied. The
absolute number of atoms is calibrated from a measurement of the
critical temperature, taking into account the effects of
interactions \cite{Ger04L}. For condensates with small condensate
fractions (less than 60\,\%), the temperature is obtained by
fitting a gaussian distribution to the thermal wings of the cloud.
The temperature is then extrapolated from the final frequency of
the rf ramp to lower temperatures for which the thermal fraction
is indiscernible \cite{Ger04,GerT03}.

\subsection{Interferometry set-up and timing}

\begin{figure}
\dasgraph{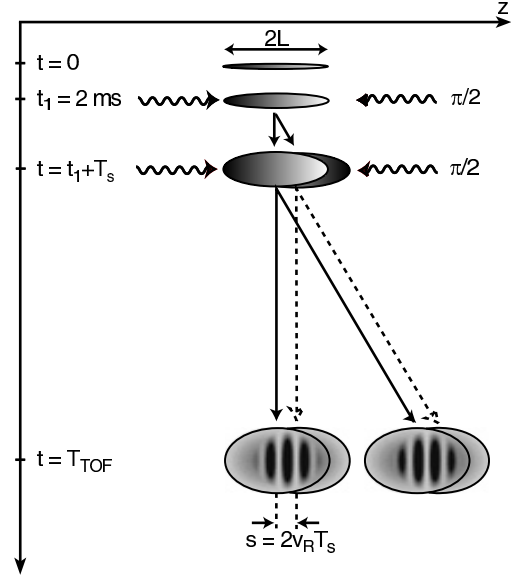} \caption{The condensate is released
from the trap at $t=0$ and has a half-length of $L$. A sequence of
two Bragg pulses is applied to generate the interferometer. Two
output ports are created ($p=0$,\ $p=2\hbar k_{\rm L}$) with
complementary fringe patterns.} \label{fig:interfero}
\end{figure}

As shown in Figure \ref{fig:interfero}, we implement the
interferometer using a sequence of two $\frac{\pi}{2}$-Bragg
pulses, which act as matter-wave beamsplitters. The set-up
consists of two laser beams counter-propagating along the
longitudinal trap axis, each of intensity
$\sim\!2\,$mW\,cm$^{-2}$, red-detuned by $\Delta=6.6\,$GHz from
the $^{87}$Rb D2 line at $\lambda=780.02$\,nm. Two acousto-optic
modulators driven by frequency synthesizers produce a small
relative detuning $\delta$, tuned to the two-photon Bragg
resonance $\delta=2\hbar k_{\rm{L}}^2/m$, with $m$ the atomic mass
and $k_{\rm L}=2\pi/\lambda$. The momentum width along $z$ of the
expanding condensate corresponds to a frequency width of 200\,Hz.
Therefore we use Bragg pulses of $\sim\!100\,\mu$s, short enough
such that the corresponding $1.6$\,kHz frequency width (full width
at half maximum) is sufficient to couple the entire condensate.
The thermal cloud surrounding the condensate has a momentum
distribution with a frequency width ranging from 12\,kHz to 60
kHz, much larger than that of the condensate. Thus, only a small
fraction of the thermal cloud is coupled by the Bragg pulses
\cite{Ger04}. By controlling the Bragg pulse length, we realize a
$\frac{\pi}{2}$-pulse which splits the condensate into a coherent
superposition of two wave\-packets with velocities differing by $2
v_{\rm{R}} = 2\hbar k_{\rm L}/m = 11.72$ mm\,s$^{-1}$, where
$v_{\rm{R}}$ is the 2-photon recoil velocity. The interferometer
sequence is illustrated in Figure \ref{fig:interfero}. The
condensate is held in the trap for at least 2\,s at the end of the
final rf evaporative-cooling ramp, to allow residual oscillations
to be damped \cite{Ric03,Shv02}. After switching off the trap, the
condensate is allowed to expand freely for 2\,ms before the first
$\frac{\pi}{2}$-pulse is applied. During this expansion the
condensate density reduces by two orders of magnitude, so
collisions between the diffracted wave\-packet and the original
condensate become negligible. During a free-evolution time
$2$\,ms\,$< T_s <10$\,ms, the two wavepackets separate to a
distance $s\,=\,2 v_{\rm R} T_s$.  The second
$\frac{\pi}{2}$-pulse completes the interferometer, and we observe
interference in each of the two output ports, which differ in
momentum by $p\,=\,2\hbar k_{\rm L}$. The condensate is imaged by
absorption perpendicular to the long axis $z$ after a 29\,ms total
time-of-flight \cite{TOF}.

For a given set of experimental conditions (condensate atom number
and temperature), the experimental correlation function is
acquired by taking a sequence of interference images with
different condensate separations $s$ ranging from $0.2L$ to
$1.2L$, varied by changing $T_s$. For smaller separations, we do
not observe enough fringes to obtain a reliable measurement of the
contrast. At the maximum value of $s$, the contrast has reduced
such that the fringes are no longer discernible above the noise.
Typical images for $0.3 L \leq s \leq 1.1 L$ are shown in Figure
\ref{fig:franges}. For each value of $s$, typically 5 images are
taken, so that a statistical average can be performed. The fringe
contrast is then measured, giving the correlation function.
Correlation functions have been obtained at various temperatures
$T$ between 100 and $230\,$nK and for condensate atom numbers
$N_0$ between $0.5\times 10^5$\ and $2.5\times 10^5$. These
conditions correspond to $0.8< T/T_{\phi}<8$, where $T_\phi=15
\hbar^2 N_0/16 m k_B T L^2$ \cite{Pet01}.

\section{Analysis of interferograms}

\subsection{Interferogram}
As shown in \cite{Pet00}, a quasi-condensate is well-described by
a fluctuating complex field $\Psi(\rho,z)= \sqrt{n(\rho,z)}
\textrm{e}^{i\Phi(\rho,z)}$, with fixed density distribution
$n(\rho,z)$ and fluctuating phase $\Phi(\rho,z)$. In the
following, $\Psi(\rho,z)$ represents the wavefunction of the
condensate after the free-fall expansion. The first Bragg pulse is
applied after 2\,ms of free expansion, at which time the density
has reduced such that interactions between the atoms are
negligible. We therefore assume that the different copies of the
condensate propagate independently. The phase distribution can be
expressed as $\Phi(\rho,z)=\alpha z^2 + \beta \rho ^2 +
\phi_{\rm{th}}(z)$, where $\phi_{\rm{th}}(z)$ represents the
thermal phase fluctuations and the quadra\-tic terms represent the
parabolic phase developed during expansion.

We now consider the interference pattern produced at one of the
output ports of the interferometer. For a separation $s$, we
obtain the atomic density distribution:
\begin{eqnarray}
n_{\rm{out}}(\rho,z) &=& \frac{1}{4} |\Psi(\rho,z-s/2)
+\Psi(\rho,z+s/2)
|^2 \nonumber\\
&=& \frac{n_+}{4}+ \frac{n_-}{4}+ \frac{ \sqrt{n_+ n_-}}{2}\cos
[\Delta \Phi(z) + \phi_{\rm{g}}], \label{equ:density}
\end{eqnarray}
with $n_\pm =n(\rho,z\pm s/2)$ and $\Delta
\Phi(z)=\Phi(z+s/2)-\Phi(z-s/2)$.  The global phase shift
$\phi_{\rm{g}}$ is due to random, uncontrolled phase shifts
between the two Bragg pulses.  The phase difference between the
two copies is $\Delta \Phi(z) =\alpha z s + \phi_{\rm{th}}(z+s/2)
- \phi_{\rm{th}}(z-s/2)$. The density of the condensate is small
when the first Bragg pulse is applied, so we neglect a small
relative velocity due to repulsion between the two copies. At the
second output port, the two condensate copies have an additional
$\pi$ relative phase shift due to the Bragg pulses, thereby
producing a complementary fringe pattern. In our data analysis,
the images of each output port are treated separately.

Since the global phase shift $\phi_{\rm{g}}$ fluctuates from shot
to shot, we cannot average over different images at the same
separation $s$. Instead, we take the contrast of each image
individually and then average the contrast.

\subsection{Analysis in Fourier Space}

\begin{figure}
\dasgraph{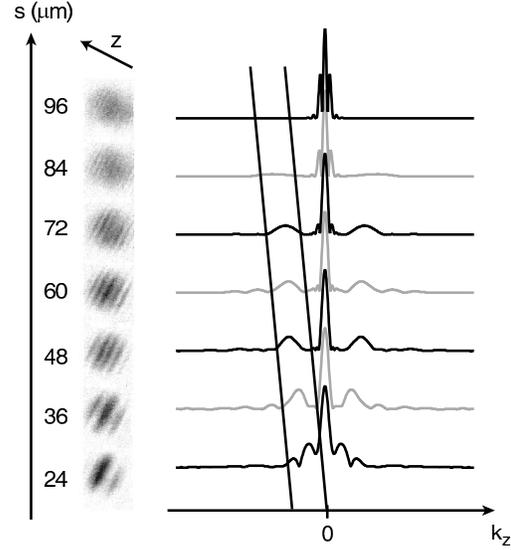} \caption{Left: One of the two output
ports of the interferometer for different separations $s$ between
the two condensate copies. The half-length of the initial
condensate is $L=85 \mu$\,m. Right: Profiles of the 2D Fourier
transform (absolute value) of each of the images on the left. The
profiles are normalized to equal central peak height (proportional
to total atom number). The position of the second peak (delimited
by the two lines) gives the spatial frequency of the fringes, and
the ratio of the height of the second peak to the central peak
gives the fringe contrast.} \label{fig:franges}
\end{figure}

\label{sec:2} The atoms are imaged by absorption along the
vertical $y$-axis, perpendicular to the long axis of the trap. The
image we obtain, rescaled to units of 2D atomic density, is the
integrated density:
\begin{eqnarray}\label{equ:image}
I(x,z) = {\rm const} \int \ud y\,n_{\rm{out}}.
\end{eqnarray}
We take the 2D Fourier transform of this image and extract its
profile along the zero radial frequency $k_x=0$ axis:
\begin{eqnarray}
\tilde{I}[0,k_z] = {\rm const} \int \ud^3 r\,n_{\rm{out}}e^{ik_z
z}.
\end{eqnarray}
Typical images and their 2D Fourier transform profiles are shown
in Figure \ref{fig:franges} for different separations $s$.  The
contrast of the fringe pattern is given by the ratio
$2\tilde{I}[0,k_0(s)]/\tilde{I}[0,0]$, where $k_0(s) \simeq\alpha
s$ is the dominant spatial frequency of the fringe pattern. The
profiles of Figure \ref{fig:franges} show clearly the increasing
spatial frequency and decreasing contrast as a function of $s$.

To extract the correlation function, we take the complex amplitude
of the Fourier peak at the spatial frequency $k_0$ of the fringes:
\begin{eqnarray}\label{equ:Itilde}
\tilde{I}[0,k_0(s)] = e^{i\phi_{\rm{g}}}\int \ud^3
r\,\sqrt{n_+n_-}e^{i\Delta\Phi_{\rm{th}}}.
\end{eqnarray}
In the absence of the global phase shifts $\phi_{\rm{g}}$, the
correlation function $C^{(1)}(s)$ would be obtained by taking the
statistical average of equation (\ref{equ:Itilde}):
\begin{eqnarray}\label{C1(s)}
C^{(1)}(s)=\langle\tilde{I}[0,k_0(s)]\rangle = \int \ud^3
r\,\langle\sqrt{n_+n_-}e^{i\Delta\Phi_{\rm{th}}}\rangle
\end{eqnarray}
which is identical to relation (\ref{equ:Cmean}). However,
$\phi_{\rm{g}}$ changes from shot to shot and prevents us from
averaging Fourier transforms directly in this way \cite{meanI}. To
eliminate this random phase, we can take the absolute value of the
Fourier transform before averaging. Thus we obtain an effective
correlation function:
\begin{eqnarray}\label{equ:C(s)}
C^{\rm{eff}}(s)=\langle\left|\tilde{I}[0,k_0]\right|\rangle =
\langle \left| \int \ud^3
r\,\sqrt{n_+n_-}e^{i\Delta\Phi_{\rm{th}}}\right| \rangle.
\end{eqnarray}
Note that although it has similar behaviour, our effective
correlation function is expected to be quantitatively different
from $C^{(1)}(s)$. Taking the absolute value of the transform
reduces the cancelling effect between the random thermal phase
shifts in the statistical average, so that the effective
correlation function $C^{\rm{eff}}(s)$ decays more slowly with $s$
than $C^{(1)}(s)$, as calculated in the next section.

\subsection{Simulation}

\begin{figure}
\dasgraph{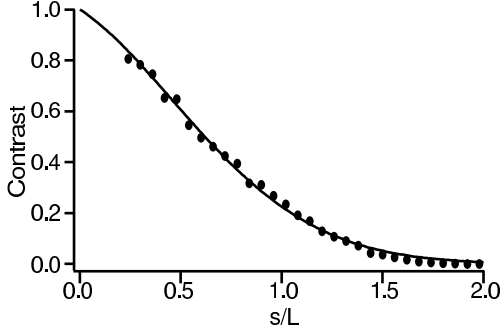} \caption{Points represent the contrast
extracted from our simulated absorption images (see text) as a
function of $s/L$, $L$ being the condensate half-length, for
$T/T_\phi = 1$. These points are fitted by the product of a
gaussian and an exponential (solid line) which we take to be our
theoretical effective correlation function $C^{\rm{eff}}(s)$.}
\label{fig:sim_points}
\end{figure}

\begin{figure}
\dasgraph{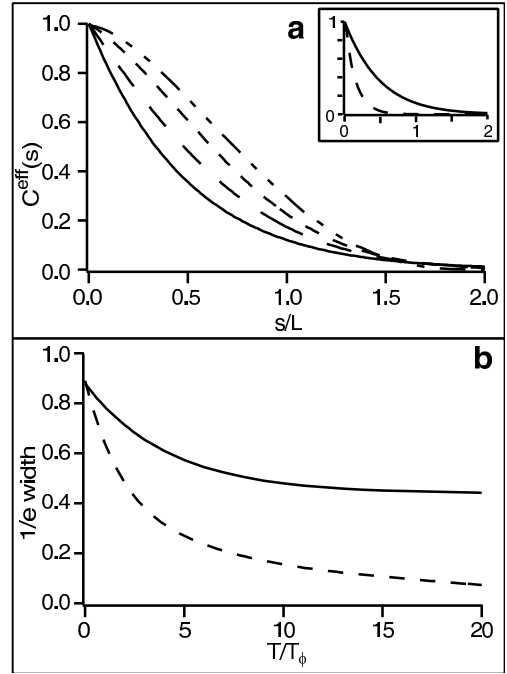} \caption{Top: Lines are simulated
effective correlation function $C^{\rm{eff}}(s)$ (described in the
text) as a function of $s/L$ for different $T/T_{\phi}$. From the
top, $T/T_{\phi} = 0, 1, 3, 10$. With increasing $T/T_{\phi}$, the
curve changes from a quasi-gaussian to an exponential and the
width decreases. Inset: Comparison of $C^{\rm{eff}}(s)$ (solid
line) and $C^{(1)}(s)$ (dashed line) at $T/T_{\phi} = 10$. Bottom:
Width at $1/e$ of $C^{\rm{eff}}(s)$ (solid line) and $C^{(1)}(s)$
(dashed line) as a function of $T/T_{\phi}$.}
\label{fig:simulation}
\end{figure}

The 1st order correlation function $C^{(1)}(s)$ can be calculated
analytically \cite{Ger03}, using the theory of \cite{Pet01} to
account for the phase fluctuations. However, it is not possible to
obtain an analytic expression for the effective correlation
function $C^{\rm{eff}}(s)$ (equation (\ref{equ:C(s)})) which we
measure. Therefore, to calculate $C^{\rm{eff}}(s)$, we first
simulate the quasi-condensate phase fluctuations following the
theory in \cite{Pet01}. The phase operator is given by:
\begin{eqnarray}
\hat\phi_{\rm{th}}(\mathbf r) = [4n(\mathbf r)]^{-1/2}\sum_j
f^+_j(\mathbf r)\hat a_j+h.c.,
\end{eqnarray}
where $\hat a_j$ is the annihilation operator of the excitation
with quantum number $j$. The solution of the Bogoliubov-de Gennes
equations for ``low energy'' excitations (with energies
$\epsilon_\nu < \hbar \omega_{\bot}$) gives the wavefunctions of
these modes:
\begin{eqnarray}
f^+_j(\mathbf r) = \sqrt{\frac{(j+2)(2j+3)gn(\mathbf
r)}{4\pi(j+1)R^2L\epsilon_j}}P^{(1,1)}_j(z/L),
\end{eqnarray}
where $P^{(1,1)}_j$ are Jacobi polynomials, $g=4\pi\hbar^2 a/m$ is
the interaction strength, $a$ is the scattering length, $R$ and
$L$ are the size of the condensate in the trap and
$\epsilon_j=\hbar\omega_z\sqrt{j(j+3)/4}$ is the energy of mode
$j$ \cite{Str98}. We assume the Thomas-Fermi approximation for the
density $n(\mathbf{r})$, taking $R$ and $L$ from fits to images
after expansion. To simulate numerically the phase fluctuations,
we replace the operators $\hat a_j$ and $\hat a^\dagger_j$ by
complex Gaussian random variables $\alpha_j$ and $\alpha^\ast_j$.
These variables have a mean value of zero and the correlation
$\langle \alpha_j \alpha^\ast_{j'} \rangle = \delta_{jj'}N_j$
where $N_j = k_B T/(\hbar \omega_z \sqrt{j(j+3)/4})$ is the
occupation number for the quasiparticle mode $j$ at a given
temperature $T$. We assume that the phase fluctuations do not
evolve on the time scale of the expansion \cite{Ger03}. We have
verified this by studying images of condensates after the same
time-of-flight, but without the interferometer pulses. In this
case, we observe a smooth density profile, with no extra features
appearing in the Fourier transform.

For a given $T/T_\phi$, $20$ condensates are generated at each
value of $s$, with $s$ ranging from $0.2 L$ to $2L$. Each
condensate is integrated over $y$ as in equation
(\ref{equ:image}). These simulated absorption images are analysed
in exactly the same way as the real experimental images. The
absolute values of the Fourier transforms of the images are
averaged, and the contrast extracted as in equation
(\ref{equ:C(s)}). The points in Figure \ref{fig:sim_points} show
the typical contrast extracted from our simulations for $T/T_\phi
= 1$. We found that these points are very well fitted by the
product of a gaussian and an exponential, for all $T/T_\phi$. We
use these fits as our theoretical effective correlation functions
$C^{\rm{eff}}(s)$.

The effective correlation function was simulated for $0\leq
T/T_\phi \leq 20$. Figure \ref{fig:simulation}a presents results
for different $T/T_\phi$. At $T=0$, $C^{\rm{eff}}(s)$ coincides
with $C^{(1)}(s)$ \cite{Ger03}. This function is simply the
integrated overlap function between the two condensates, and is
approximately a gaussian function of the separation $s$
\cite{Zam00}. As $T/T_{\phi}$ increases, the width of the function
decreases and its form gradually becomes exponential. In Figure
\ref{fig:simulation}b we plot the $1/e$ widths of the simulated
$C^{\rm{eff}}(s)$ functions as a function of $T/T_\phi$. For
comparison we show also the width of $C^{(1)}(s)$ \cite{Ger03}
which decreases much faster with $T/T_{\phi}$.

\section{Experimental results}

\begin{figure}
\dasgraph{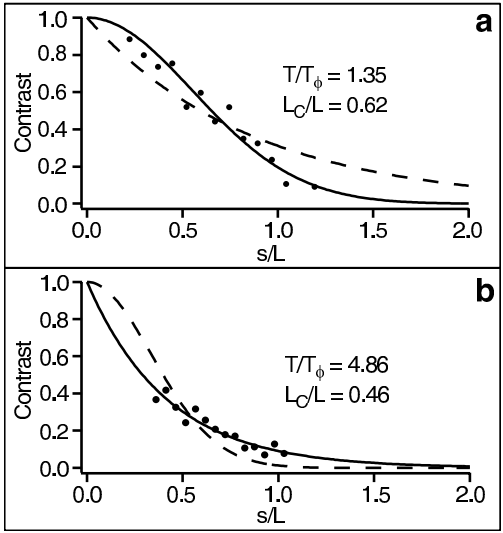} \caption{Example of two experimental
effective correlation profiles (points) as a function of $s/L$
showing clearly the change in shape and width with $T/T_{\phi}$.
Each point is averaged over 5 condensates, each with 2 output port
images. Top: At $T/T_{\phi}=1.35$, the effective correlation curve
is fitted by a gaussian (solid line), with $1/e$ width
$L_{\rm{C}}/L=0.62$. The fit to an exponential (dashed line) is
shown for comparison. Bottom: At $T/T_{\phi}=4.86$, the effective
correlation curve is fitted better by an exponential (solid line),
with $1/e$ width $L_{\rm{C}}/L=0.46$. The fit to a gaussian
(dashed line) is shown for comparison.} \label{fig:contraste}
\end{figure}

Figure \ref{fig:contraste} shows two examples of effective
correlation curves measured using our interferometer and analysed
as described above. The points shown in Figure
\ref{fig:contraste}a were obtained using a magnetic trap with an
aspect ratio of 45, and with an atom number and temperature
corresponding to $T/T_\phi=1.35$, that is for small-amplitude
phase fluctuations. The points in Figure \ref{fig:contraste}b were
obtained using a trapping aspect ratio of 100, and with
$T/T_\phi=4.86$. The contrast is plotted as function of $s/L$,
obtained from a fit to a truncated parabola. Each point
corresponds to an average over 5 condensates. The difference in
the range of $s/L$ explored is due to different expansion dynamics
after release from the two different traps. In the more tightly
confined trap (Figure \ref{fig:contraste}b), the axial expansion
is much slower, and thus the fringe spacing decreases more slowly
with $s$. At the smallest values of $s$, it is therefore
impossible to measure the contrast reliably since we do not
observe enough fringes. We can extract information about the phase
fluctuations from both the shape and the width of these effective
correlation functions.

\subsection{Shape of the effective correlation functions}
First, we observe qualitatively that the shape of the effective
correlation functions $C^{\rm{eff}}(s)$ changes as $T/T_{\phi}$
increases. For small $T/T_{\phi}$, as in Figure
\ref{fig:contraste}a, the curves are clearly gaussian, as shown by
the fit in the figure. As we increase $T/T_\phi$, the profiles
become rapidly exponential.  We see in Figure \ref{fig:contraste}b
that at $T/T_\phi=4.86$, the curve is already better fitted by an
exponential than a gaussian.  At intermediate values of
$T/T_\phi$, we can use the product of a gaussian and an
exponential to fit a smooth curve through the data.  The
contribution of the exponential increases rapidly in importance at
finite $T$, in agreement with the simulation, reflecting the
increasing amplitude of the phase fluctuations with $T/T_\phi$.

\subsection{Comparison of coherence length \boldmath${L_{\rm{C}}}$}

\begin{figure}
\dasgraph{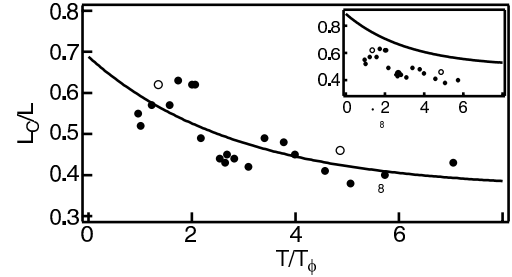} \caption{$L_{\rm{C}}/L$ as a function
of $T/T_\phi$. The solid curve is a fit using the function
described in the text. Empty circles correspond to the two
experimental correlation profiles of Figure \ref{fig:contraste}.
Inset: The solid curve is the simulated effective correlation
function of Figure \ref{fig:simulation}b.}
\label{fig:Graphe_TTphi}
\end{figure}

In order to extract quantitative information from the effective
correlation functions, we define a coherence length $L_{\rm{C}}$,
equal to the $1/e$ width of the effective correlation curve
$C^{\rm{eff}}(s)$. We then use this parameter to compare the
widths of the measured and simulated effective correlation
functions. A smooth curve is fitted through the data (using the
product of a gaussian and an exponential) and the $1/e$ width
extracted. Although the thermal cloud plays no role in the
interference pattern we observe, it appears behind the condensates
in the $p = 0$ output of the interferometer and thus reduces the
measured contrast. Independent measurements of the thermal
fraction (between 60\,\% and 80\,\% for this experiment) allow us
to renormalize the experimental effective correlation functions to
take account of this effect. When fitting the renormalized curves,
we then fix the value at $s/L=0$ to unity.

In Figure \ref{fig:Graphe_TTphi}, we plot $L_{\rm{C}}/L$ as a
function of $T/T_{\phi}$. Importantly, we see that the coherence
length varies smoothly as a function of $T/T_\phi$, even when the
temperature is close to $T_\phi$. This is what we should expect
since  $T_{\phi}$ is simply a characteristic temperature, defined
as that at which the mean square fluctuations of the phase
difference between two points separated by a distance $L$ is equal
to 1. Therefore it should be borne in mind that even condensates
at temperatures below $T_\phi$ are not necessarily fully coherent.

The inset of Figure \ref{fig:Graphe_TTphi} compares the measured
coherence widths with the results of the simulation (Figure
\ref{fig:simulation}b). Despite the offset between the two curves,
the trend of the data follows very well that of the simulation. In
Figure \ref{fig:Graphe_TTphi} the experimental data is fitted by a
curve $A+B\exp\left(-b T /T\phi\right)$, giving $b=0.35$ and
$B=0.32$.  This is in reasonable agreement with the simulation,
for which the fit yields $b=0.32$ and $B=0.39$. In the following
discussion we consider possible explanations for the observed
reduction in contrast.

\subsection{Discussion}
As shown in the inset of Figure \ref{fig:Graphe_TTphi}, the
measured coherence length is offset by about $20\%$ from the
results of the simulation at $T/T_{\phi}=0$. In order to eliminate
various possible causes of this discrepancy, and to understand
better the limitations of our experiment, we have performed
several tests.

Since the accuracy of our experiment relies on the comparison of
fringe contrast at different spatial frequencies, it is important
to take great care in setting up and characterising the imaging
system. Therefore we measured the modulation transfer function
(MTF) (see {\it e.g.} \cite{Goodman}) of our complete imaging
system \textit{in situ}, using a USAF1940 resolution target
engraved with 3-bar square wave patterns of spatial frequency
4--200 lines/mm, covering the range of spatial frequencies
observed in our interference experiment. By Fourier transforming
images of different regions of the target, we were able to compare
the magnitude of the different Fourier components in the image
with those of the target pattern, thereby obtaining the MTF of our
system.  We found that the MTF is approximately linear, falling
from 1 at zero spatial frequency to 0 at $118\,$lines/mm. This
resolution limit at $8.5\,\mu$m is in agreement with earlier
characterisations of the system \cite{RicT03}. The shape of the
MTF is due almost entirely to the CCD camera \cite{PixelFly}, and
is surprisingly significant at fringe spacings much greater than
the effective pixel size of $2.5\,\mu$m. All contrast measurements
were corrected by this MTF. In fact, for the data obtained using
the second trapping aspect ratio, the axial expansion (and thus
the phase difference developed) was sufficiently small that the
maximum observed fringe spatial frequency was $16\,$lines/mm and
therefore the MTF correction had a negligible effect on the
experimental effective correlation functions. In the first set of
data, where the maximum fringe frequency was $38\,$lines/mm (a
fringe spacing of $25\,\mu$m), the correction was more
significant, changing the width of the curves by typically 10\%,
though still leaving a 20\% discrepancy with the simulation.

We also considered the error introduced by a small focus\-sing
error. The imaging system is focussed onto the condensate to
within $\pm 0.2\,$mm by minimising the imaged size of a small
condensate as a function of the objective lens position. However,
by varying the time-of-flight used, we may have introduced an
error of up to $\pm 1\,$mm.  Such an error could also be
introduced by small changes to the residual magnetic fields, which
lead to changes in the release velocity of the condensate when the
magnetic trap is switched off.  We measured effective correlation
curves for different foci of the imaging system, but found that
the width of the effective correlation curve changed by less than
the existing spread in the points.

Other possible sources of error, such as the alignment of the
imaging beam with respect to the condensates' fringes and correct
background subtraction have also been eliminated.

There remains in our experiment an unexplained pheno\-menon
regarding the distribution of the fringe pattern.  We expect to
see fringes only in the region where the two condensates overlap.
However, it can be seen in the images of Figure \ref{fig:franges}
that the fringes extend to the edges of each condensate. Moreover,
these ``extra'' fringes have the same spatial frequency and phase
as the central fringes.  Although it is possible that a small
fraction of the thermal cloud is coupled by the Bragg beams,
interference fringes produced in this way \cite{KetterleThermal}
would have a much smaller fringe spacing, less than $6\,\mu$m.
More importantly, the contributions from different parts of the
original thermal cloud, whose width is $\sim\!100\,\mu$m, would
sum incoherently to wash out the fringe pattern. It is more likely
that these extra fringes arise as a result of interactions during
the application of the Bragg pulses, but better modelling is still
needed before evaluating whether their presence should increase or
decrease the overall measured contrast.

\section{Conclusion}
\label{sec:1}
We have demonstrated a new type of matter-wave interferometry
using Bragg beam-splitter pulses and Fourier space analysis. Our
results show that the expected shape of the correlation functions
changes from a gaussian-like shape to an exponential-like shape
when the amplitude of phase fluctuations is increased. The
coherence length of elongated condensates varies smoothly at
temperatures close to $T_\phi$, as predicted by theory. This
highlights the fact that the characteristic phase temperature
$T_\phi$ does not indicate a transition to full phase coherence,
but rather that condensates exhibit phase fluctuations at all
finite temperatures, albeit of small amplitude. This may place
constraints on the trapping geometries which can be used for
creating measurement devices based on the phase coherence of
condensates.

\section{Acknowledgements}
We would like to acknowledge support from IXSEA-OCEANO (M.H.),
from the BEC2000+ programme and the Marie Curie Fellowships
programme of the European Union (J.R.), from IXCORE (F.G.) and
from the Chateaubriand Program, CNRS, and NSERC (J.H.T.). This
work was supported by D\'{e}l\'{e}\-gation G\'{e}n\'{e}rale de
l'Armement, the European Union (grants IST-2001-38863 and
MRTN-CT-2003-505032) and INTAS (Contract No. 211-855).


\begin{thebibliography}{}
\bibitem{Bragg} Note that the Bragg momentum spectroscopy method
does not suffer from this problem.
\bibitem{TOF} Our cell allows maximum 25\,ms free-fall. We increase the time-of-flight by applying, after the Bragg pulses, a magnetic field gradient opposed
to the gravity field.
\bibitem{meanI} This would be equivalent to averaging the original images: the
fringe contrast would be averaged out to the non-interfering
density profile.
\end{thebibliography}
\end{document}